\documentclass[amssymb,amsmath,amsthm,superscriptaddress,
reprint,aps,pra]{revtex4-2}
\usepackage{graphicx}	%
\usepackage{dcolumn}	%
\usepackage{bm,bbm}		%
\usepackage{upgreek}
\usepackage{braket}
\usepackage{xcolor}		 %
\usepackage[many]{tcolorbox}
\usepackage{multirow}
\usepackage{subcaption}

\usepackage[bookmarks=false]{hyperref}	%
\hypersetup{colorlinks=true,citecolor=blue,linkcolor=magenta,
	urlcolor=cyan,pdfstartview=FitH,bookmarksopen=true}
\makeatother

\usepackage{csquotes}
\MakeOuterQuote{"}

\usepackage{soul}	%
\soulregister\cite7
\soulregister\ref7
\soulregister\citep7
\soulregister\citet7
\soulregister\pageref7

\newcommand{\kt}[1]{\vert #1 \rangle}

\newcommand{\AffiBAQIS}{\affiliation{Beijing Academy of Quantum Information Sciences, Beijing 100193, China}}
\newcommand{\AffiBUPT}{\affiliation{School of Science and State Key Laboratory of Information Photonics and Optical Communications, Beijing University of Posts and Telecommunications, Beijing 100876, China}}
\newcommand{\AffiULB}{\affiliation{Laboratoire d’Information Quantique, Université Libre de Bruxelles, Brussels 1050, Belgium}}

\begin{document}
	
	\title{Transmitter-device-independent quantum key distribution}
	
	\author{Qiang Zeng}	\email{zengqiang@baqis.ac.cn}		
	\AffiBAQIS
	
	\author{Abhishek Mishra}
	\AffiULB
	
	\author{Haoyang Wang}
	\AffiBAQIS
	\AffiBUPT
	
	\author{Zhiliang Yuan}	%
	\AffiBAQIS
	
	\date{\today}
	
	\begin{abstract}
		Transmitter-side device dependence is a longstanding yet implicit problem in quantum key distribution (QKD), in contrast to the thoroughly addressed measurement-side case. 
		Quantum steering, which intrinsically distinguishes the roles of sender and receiver in entanglement certification, offers a natural route to lifting trust assumptions on the transmitter.
		Existing one-sided device-independent (1sDI) protocols that exploit quantum steering as the security resource typically operate in a tripartite architecture, whereas adapting them to a transmitter-receiver architecture can introduce a signaling loophole.
		Here we conceptually formalize the transmitter-device-independence on the basis of faithful quantum steering and propose a practical transmitter-device-independent (TDI) QKD protocol within the 1sDI trust model.
		The TDI scheme operationalizes the implicit no-signaling condition by enforcing a causal isolation between the input information and the transmitted quantum states, thereby eliminating the associated transmitter-side device vulnerabilities.
		In a proof-of-principle experiment, we validate the feasibility of the TDI protocol, achieving a key rate in the asymptotic limit of 410~bps over 27~km spool fiber. 
		By delivering TDI security while retaining strong loss tolerance, our approach helps bridge the gap between security and practicality for real-world QKD deployments.
	\end{abstract}
	
	\maketitle
	
	\begin{figure*}[ht]
		\centering
		\includegraphics[width=1.7\columnwidth]{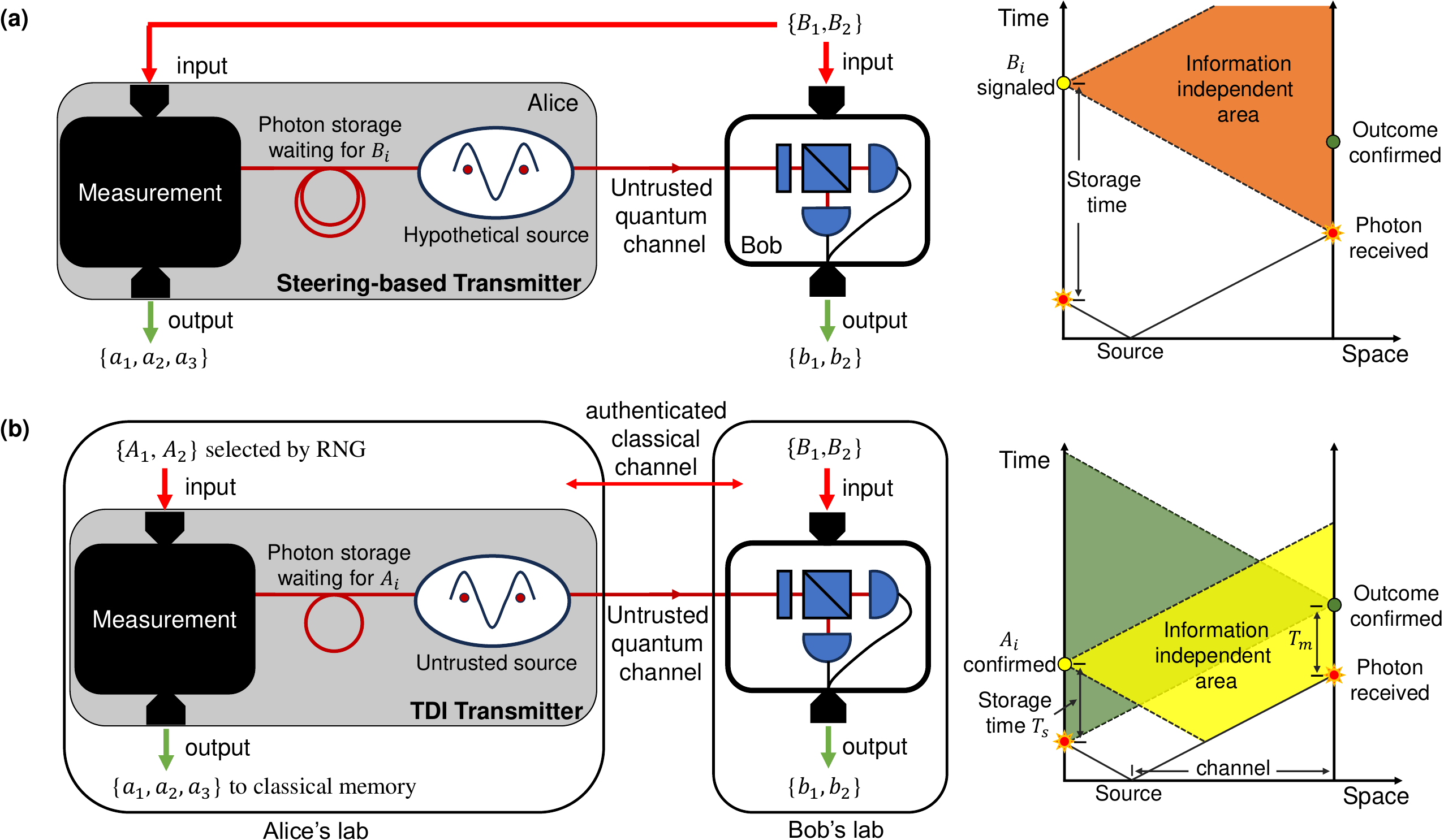}
		\caption{(a) Faithful execution of quantum steering.
			Left panel: steering certification schematic.
			Alice and Bob denote the abstract parties.
			The inputs of Alice are determined by Bob.
			Right panel: space-time diagram of steering framework.
			The photon for Alice's measurement is stored until Bob's input information is signaled, and 	the storage time is distance-dependent.
			(b) Left panel: TDI-QKD schematic.
			Alice and Bob each operate within an isolated laboratory, represented schematically by an enclosed box.
			Both parties are equipped with trusted random number generators (RNG) and trusted classical memories, while the internal mechanism of Alice’s transmitter is left uncharacterized and Bob’s receiver is well characterized.
			The authenticated classical channel opens only when the quantum channel ceases to work, and only the information for security analysis is permitted to exchange.
			Right panel: space-time diagram of TDI protocol.
			The signaling loophole is closed because Alice releases her input choice only in the information independent area of Bob's outcome (the green triangle).
			The photon for Alice is stored until her information is released.
			The storage time $T_s$ is determined only by the measurement interval $T_m$ of Bob's devices, that is, $T_s > T_m$, being irrelevant to the channel length from the transmitter to the receiver.
		}\label{f1}
	\end{figure*}
	
	\section{Introduction}
	Secret sharing is a cornerstone of modern information security. 
	In contrast to classical cryptography, which relies on computational assumptions, quantum key distribution (QKD) derives security from fundamental physical principles, whereby any eavesdropping attempt necessarily introduces detectable disturbances~\cite{BB842014,loDecoyStateQuantum2005,wangBeatingPhotonNumberSplittingAttack2005,xu2020RMP}.
	In practice, deviations of the prepared states and measurements from their ideal designs can be exploited to compromise security, so precise device calibration is required. 
	In extreme cases, an adversary may actively manipulate the devices to report false outcomes, undermining the security analysis and compromising real-world QKD systems~\cite{gisin2006Trojanhorse,lydersenHackingCommercialQuantum2010}. 

	The ultimate solution is device-independent (DI) QKD~\cite{mayersQuantumCryptographyImperfect1998}, which treats the devices as black boxes and certifies security directly from observed nonlocal statistics~\cite{Primaatmaja2023securityofdevice,zapateroAdvancesDeviceindependentQuantum2023,ghoreishiFutureSecureCommunications2025}. 
	Although this offers the highest level of security, a fully DI implementation 	imposes exceptionally stringent experimental conditions to close physical loopholes such as the detection loophole~\cite{pearleHiddenVariableExampleBased1970,gerhardtExperimentallyFakingViolation2011}, and its attainable key rate and communication distance remain limited~\cite{nadlingerExperimentalQuantumKey2022,zhangDeviceindependentQuantumKey2022,liuPhotonicDemonstrationDeviceIndependent2022,DIQKD2026_100km}.

	Measurement-device-independent (MDI) QKD~\cite{MDI2012} exploits two-photon interference to postselect %
	entanglement, %
	which can be regarded as a time-reversed version of entanglement-based QKD.
	By removing all trust in the measurement device, MDI-QKD eliminates all detector side channels while retaining practicality, and later variants that exploit single-photon~\cite{lucamariniOvercomingRateDistance2018} or asynchronous two-photon interference~\cite{zengModepairingQuantumKey2022,xieBreakingRateLossBound2022}  pushed the achievable distance substantially further~\cite{zhou2023TwinfieldQuantum,zhouExperimentalQuantumCommunication2023,liuExperimentalTwinFieldQuantum2023}.
	However, all prepare-and-measure (P\&M) protocols~\cite{ioannouReceiverDeviceIndependentQuantumKey2022a,ioannouReceiverdeviceindependentQuantumKey2022,wangFullyPassiveQuantumKey2023}, including MDI-QKD, remain susceptible to source side-channel vulnerabilities, which arise both from imperfections in physical devices~\cite{tamakiDecoystateQuantumKey2016,luHackingMeasurementdeviceindependentQuantum2023} and from uncharacterized state dimensions~\cite{pironio2009DeviceindependentQuantum}.
	To date, the complementary notion of transmitter-device-independence, i.e., the natural counterpart to MDI, has yet to be formalized and its significance remains underappreciated.
	
	A transmitter-device-independent (TDI) protocol is particularly well suited to network architectures in which mobile users equipped with less secure terminals communicate with trusted central servers.
	Transmitter-device-independence can be naturally established by certifying nonlocal correlations in an asymmetric model, thereby placing the problem within the framework of quantum steering~\cite{wiseman2007Steering,uolaQuantumSteering2020}.
	Indeed, quantum steering has been harnessed as the resource to implement one-sided device-independent (1sDI) QKD.
	The concept was first implied by entropic uncertainty relations~\cite{tomamichelUncertaintyRelationSmooth2011} and later formalized through operational 1sDI-QKD protocols~\cite{tomamichelTightFinitekeyAnalysis2012}. 
	Experimental realizations were first achieved in continuous-variable systems~\cite{gehringImplementationContinuousvariableQuantum2015,walkExperimentalDemonstrationGaussian2016}, but their security requirements remain stringent and consequently limit the communication distance. 
	A discrete-variable 1sDI-QKD protocol has also been proposed~\cite{branciardOnesidedDeviceindependentQuantum2012}, but remains experimentally unrealized owing to its marginal relaxation on the experimental requirement and its vulnerability to coherent attacks.
	Recent numerical techniques~\cite{masiniOnesidedDIQKDSecure2026,brownComputingConditionalEntropies2021,brownDeviceindependentLowerBounds2024} suggest that a 1sDI-QKD protocol could reach a secure communication distance comparable to that of conventional QKD systems, but the experimental validation and performance benchmarking are still lacking.

	However, 1sDI-QKD does not constitute a faithful execution of quantum steering, as it decouples the causality of the measurement choices.
	In this work, we formalize transmitter-device-independence on the basis of faithful quantum steering and identify a signaling loophole in existing 1sDI-QKD implementations that adopt a quantum steering architecture.
	Building on the 1sDI trust model, we propose a TDI protocol that closes the signaling loophole by enforcing information isolation between the input information of the sender and the transmitted quantum states to the receiver.
	Unlike the strict relativistic causal arrangement employed in quantum steering, the TDI protocol operates under a causal condition tailored to the cryptographic setting, thereby enabling a practical implementation of TDI quantum cryptography.
	Finally, we experimentally validate the protocol in a fully photonic platform based on polarization encoding, paving the way for practical implementations of TDI-QKD.

	\section{Transmitter-device-independence}
	In the P\&M scenario, the transmitter as a single device produces the classical outcome and the accompanying quantum state, and the information security relies on the eavesdropper's uncertainty of the key bits given the emitted quantum states.
	This uncertainty is not guaranteed if the transmitter is untrusted as one cannot rule out the case that a preprogrammed classical computer inside the transmitter simulates the statistics.
	While the internal mechanism of a single device cannot be self-tested~\cite{supicSelftestingQuantumSystems2020}, having a multipartite configuration can, in principle, validate the nonclassical randomness.
	
	The appropriate quantum scenario to model an untrusted transmitter and a trusted receiver is quantum steering. 
	A defining question in this scenario is whether a collection of quantum states admits local hidden state (LHS) models~\cite{pusey2013Negativity, zeng2022reliable,zeng2022OnewayEinsteinPodolskyRosen}. 
	As illustrated in the left panel of Fig.~\ref{f1}(a), a hypothetical source at Alice's side distributes entangled photon pairs to the two parties, where Alice is untrusted and Bob is trusted.
	Upon receiving the photons, Bob randomly selects his measurements from the set $\{B_1, B_2\}$ and obtains dichotomic outcomes denoted by $b_1$ or $b_2$.
	Bob's measurement choices are signaled to Alice as her inputs only after Bob has received the photon to be measured, thus guaranteeing the no-signaling condition between Alice's operations and Bob's states.
	Alice obtains outcomes in a set $\{a_1, a_2, a_3\}$, where $a_3$ denotes a failure
	to produce a detection event, typically corresponding to photon loss in practice.
	This protocol implements quantum steering certification: observing a violation of the steering inequalities guarantees a nonclassical (i.e., refuting the LHS models) state distribution from Alice to Bob~\cite{zeng2018}.
	That is to say, treating Alice as the transmitter, certifying 
	steering nonlocality establishes transmitter-device-independence~\cite{zeng2025}.
	
	It should be noted that, to date, no QKD protocol corresponding to a faithful implementation of a quantum steering test has been proposed.
	The primary challenge lies in closing the locality loophole, which requires enforcing a relativistic causal ordering of the measurement operations to justify the no-signaling condition.
	This necessitates storing the photon on the untrusted party until a signal is received from the trusted party.
	As depicted in the right panel of Fig.~\ref{f1}(a), the required storage time scales proportionally with the transmission distance.
	In the fully photonic loophole-free demonstration~\cite{wittmannLoopholefreeEinsteinPodolsky2012}, the storage is realized through optical delay using a fiber spool.
	However, this introduces a distance-dependent photon loss, ultimately limiting the achievable communication distance to the trusted party.
	Alternatively, one would consider a long-lived quantum system [trapped ions, atoms, nitrogen-vacancy centers] to produce the entangled states and use Bob's signal to herald a successful detection.
	However, such systems are limited in clock rate, and their experimental complexity is comparable to that of DI-QKD, which diminishes the practical appeal of steering-based QKD.
	
	The 1sDI-QKD protocols adopt an asymmetric trust model similar to that of the quantum steering.
	However, unlike quantum steering, it is formulated as a direct modification of the fully DI framework and therefore comprises three distinct entities: the source and the two communicating parties---note that the two parties are assumed to have secure input-output information independent of their respective devices.
	It commits trust assumption on one party's device to ease the requirement of closing loopholes on the other side.
	The 1sDI-QKD protocols, although exploiting quantum steering as the security resource, do not correspond to a faithful execution of quantum steering, wherein both parties randomly and independently select inputs without designating the sender or receiver.
	In existing 1sDI-QKD demonstrations~\cite{gehringImplementationContinuousvariableQuantum2015,walkExperimentalDemonstrationGaussian2016}, Alice's detection and the source are effectively treated as a single party following the quantum steering architecture.
	We do not claim that the mentioned experiments have been demonstrated to be insecure, nor do we claim that their reported experimental statistics constitute evidence of a successful attack. 
	However, the no-signaling condition can be compromised as the inputs $A_i$ and $B_i$ are causally decoupled, which consequently undermines the system's security.
	For instance, once Alice's input information becomes accessible, a malicious local source embedded within the device could exploit the distribution channel to send predetermined states in reverse, hijacking the states intended for Bob thus fabricating nonlocal statistics.
	More generally, without proper information isolation, Alice's device and the untrusted source effectively function as a single entity.
	We refer to this potential vulnerability as the \textit{signaling loophole}, which constitutes a specific manifestation of the locality loophole.
	See Appendix for a concrete attack model.
	
	To eliminate the signaling loophole arising in adopting a transmitter-receiver architecture in the 1sDI framework, we propose a protocol in which a causal isolation is enforced.
	In specific, Alice’s input information is supplied to the transmitter after a sufficient time delay following the generation of the entangled photon pair, and the photon destined for Alice is stored until she releases her input.
	Consequently, as illustrated in the right panel of Fig.~\ref{f1}(b), Bob’s outcome is causally independent of, and hence informationally isolated from, Alice’s input.
	The TDI transmitter therefore comprises an entanglement source, a photon-storage module, and a measurement device, as illustrated in the left panel of Fig.~\ref{f1}(b), deployed at Alice’s isolated laboratory.
	Despite its defined functional structure, the transmitter is treated as a single untrusted device, without imposing any assumptions on or characterization of its internal workings.
	For instance, the photon storage is only one possible physical implementation for realizing the required causal ordering; the security condition is imposed on the resulting causal-isolation condition in the information flow, rather than on the internal operation or even the existence of the storage module.
	On the other hand, Bob equipped with trusted devices acts as the receiver.
	It is worth emphasizing that Alice controls the timing at which her input information is generated and released, using a trusted random-number generator, independently of the behavior of the transmitter. 
	Thus, the timing of Alice’s input can be trusted without requiring any assumptions about the transmitter’s internal workings or introducing any additional trust assumption into the security model.

	Unlike faithful quantum steering, the delay/storage time is determined solely by the measurement time interval of the receiver's devices, therefore enabling a viable channel length to Bob.
	In this sense, the signaling loophole in the context of cryptography can be viewed as a weakened form of the locality loophole in the context of nonlocality certification of Nature.
	Specifically, it constrains the signaling carrying Alice's input $A_i$ to propagate to Bob no faster than the transmission speed of the quantum channel, e.g., the speed of light in an optical fiber.
	This constraint is justified by assuming that Alice’s and Bob’s laboratories are physically isolated, such that any unintended signaling channels between the two parties, other than the legitimate quantum channel, are excluded.
	To summarize, in contrast to conventional steering-based protocols, the proposed TDI protocol operates under a tailored causal-isolation condition, in which Alice’s input is generated by a trusted randomness source and the speed of any signaling associated with her input is subject to a prescribed propagation-speed constraint.
	Under this condition, the protocol excludes LHS models, thereby establishing TDI security in a practical cryptographic setting.

	\section{TDI-QKD protocol}
	The proposed TDI protocol rests on the 1sDI trust model, that is, beyond the minimal set of assumptions:
	(i) the validity and completeness of quantum theory; 
	(ii) the availability of trusted local randomness for input generation by the communication parties; 
	(iii) isolation of the input-output information from adversaries;
	and 
	(iv) an authenticated classical channel and trusted classical postprocessing,  %
	it additionally assumes that 
	(v) the measurement devices at the receiver are precalibrated and certified---especially, the no-click outcomes are independent of the input choice, known as fair-sampling assumption in the context of Bell test~\cite{brunner2014Bell}.
	
	Now we describe the detailed procedures of TDI-QKD protocol.
	\begin{tcolorbox}[breakable, colback=gray!10, colframe=black, title=TDI-QKD protocol]
		\begin{enumerate}
			
			\item \textbf{Preparation.}
			Bob certifies the parameters of his measurement apparatuses including the measurement interval. 
			Alice and Bob certify the total transmission time of the channel, synchronize their local time, and reconcile an input clock for the system.
			They verify the timing and causal-isolation condition in test runs, that is, the two events releasing $A_i$ and registering $b_i$ are informationally independent.
			
			\item \textbf{Statistics accumulation.}
			The transmitter is adapted to the input clock and distributes photons to Bob via the quantum channel.
			Alice and Bob randomly select inputs and accumulate outcomes including the associated time tags.
			
			\item \textbf{Security analysis.}
			Upon sufficient accumulation, the quantum channel ceases to work. 
			Bob announces his choice of inputs via public classical channel and reconciles with Alice which of those rounds are for security analysis.
			Alice receives Bob’s time-tagged outcomes of the reconciled rounds, compares these time tags against her own records, matches those abiding by the causal-isolation condition, and then computes the conditional probability distribution $p(a,b|A,B)$.
			
			\item \textbf{Key generation.}
			Alice computes the key rate based on $p(a,b|A,B)$.
			If the key rate is positive, they exchange necessary information via authenticated classical channel and perform error correction and privacy amplification on their respective key-generation outcomes, otherwise they abort the protocol.
			
		\end{enumerate}
	\end{tcolorbox}
	
	For key-rate analysis, we take $A_1$ and $B_1$ as the key-generation combination.
	As the TDI-QKD protocol satisfies the no-signaling condition, we are able to evaluate the key rate in the asymptotic regime under one-way classical postprocessing from Alice to Bob using the Devetak-Winter bound:
	\begin{equation}
		r_{\infty} \ge H(b|E,B_1)-H(b|a,A_1,B_1),
	\end{equation}
	where $H(b|E,B_1)$ represents the conditional von Neumann entropies of Bob's outcomes by measurement $B_1$ given Eve’s information, and $H(b|a,A_1,B_1)$ represents the conditional Shannon entropies of outcomes of $B_1$ given Alice's outcome by $A_1$.
	These two quantities quantify the extent to which Eve and Alice, respectively, can reproduce Bob's outcomes.
	The quantity $H(b|a,A_1,B_1)$ can be directly estimated from the observed input-output statistics associated with the measurement settings $A_1$ and $B_1$. 
	In contrast, estimating $H(b|E,B_1)$ is considerably more challenging. 
	To this end, we employ a recently developed numerical technique to perform the security analysis (see Appendix).

	\begin{figure*}[t]
		\centering
		\includegraphics[width=1.7\columnwidth]{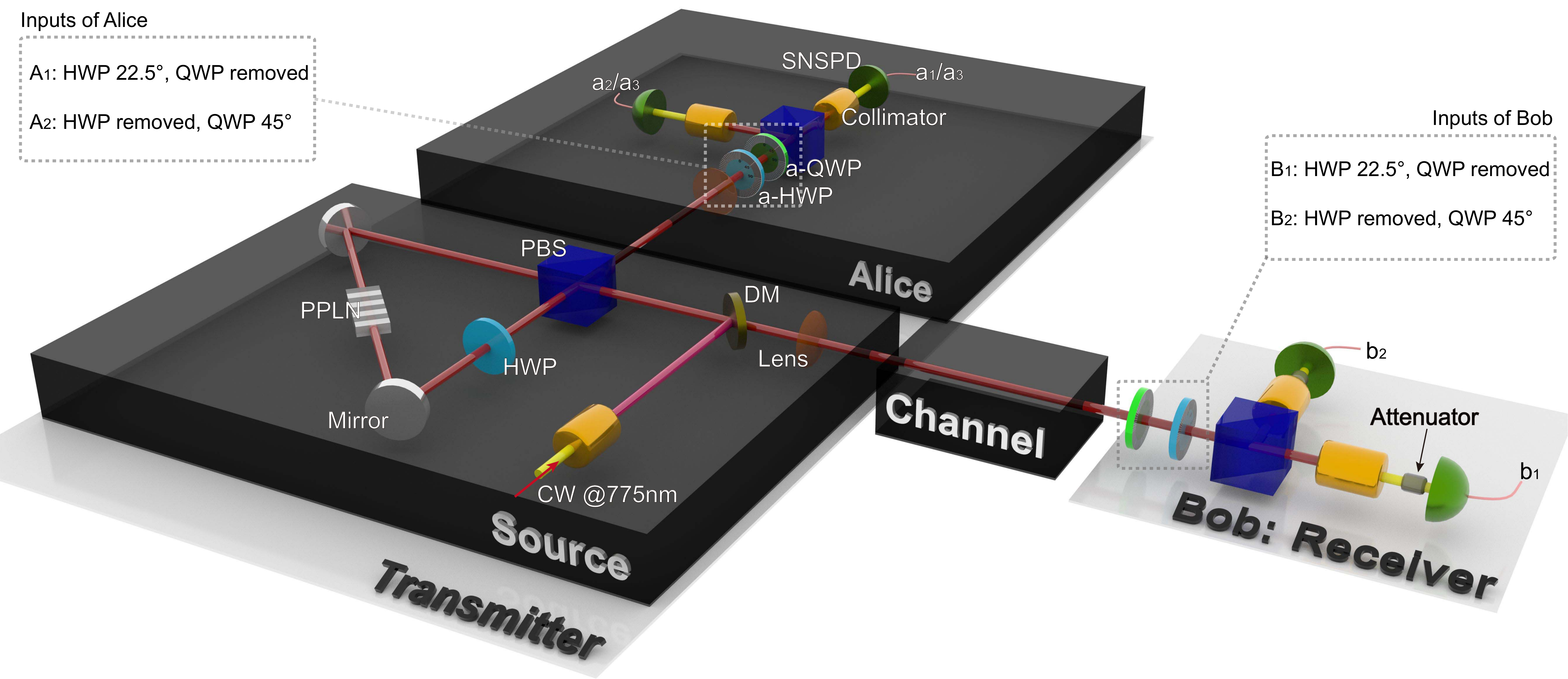}
		\caption{Experimental setup. 
			The setup of transmitter is contained in three virtual black boxes indicating the internal mechanisms are not postulated.
			Abbreviations: a-HWP, active half wave plate; a-QWP, active quarter wave plate; DM, dichroic mirror; PBS, polarization beam splitter; NBF, narrowband filter; PC, polarization controller; SNSPD, superconducting nanowire single-photon detector.
		}\label{f2}
	\end{figure*}

	\section{Experiment and results}
	We demonstrate a proof-of-principle experiment to validate the feasibility of the protocol in a fully photonic system.
	As illustrated in Fig.~\ref{f2}, the transmitter comprises three functional components, the entanglement source, the photon storage realized by fiber delay, and the measurement module.
	The measurement interval of the detector is calibrated to be 50~ns, for which we employ a 30-m fiber corresponding to a 150~ns delay in the storage module to enforce the causal isolation.
	The experimental details are provided in Appendix.
	
	The feasibility of the protocol is largely determined by the performance of the transmitter device.
	In particular, the specific internal mechanism of Alice's device is not postulated, which means the no-click outcomes cannot be safely discarded therefore posing a detection-efficiency threshold on her device.
	The general detection model for Alice and Bob is described by an eight-element positive-operator-valued-measure (POVM) set including four different click events, involving two experimental parameters: the detection efficiency $\eta$ and the dark count probability $p_{\text{D}}$.
	A detailed description of the detection model is provided in Appendix.
	
\begin{figure}[h!]
	\centering
	\includegraphics[width=\columnwidth]{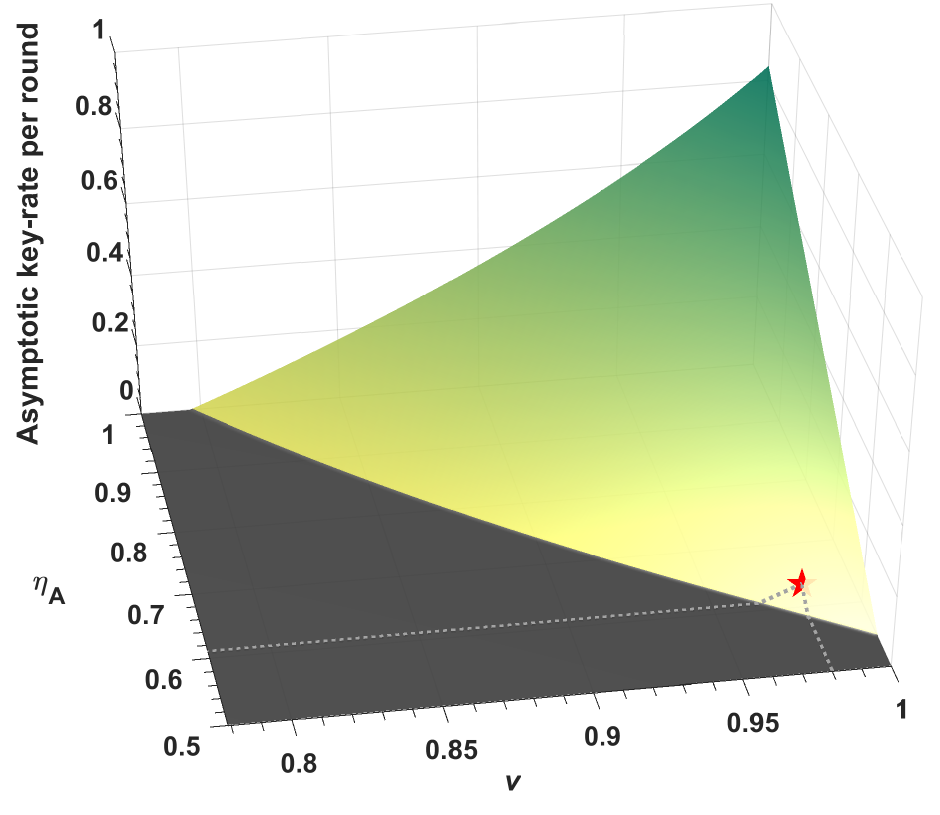}
	\caption{Key-rate bound by the performance of the transmitter.
		Parameters $\eta_A$ and $v$ are the heralding efficiency of Alice and the visibility of the entanglement state, respectively.
		The simulation adopts $\eta_B=18\%$, $p_{\text{D}}^A=5 \times 10^{-4}$, and $p_{\text{D}}^B=2 \times 10^{-3}$.
		The red pentagram denotes our experimental data point, corresponding to an $\eta_A$ of 61\% and a $v$ of 98\%.
	}\label{f3}
\end{figure}

	In the experiment, the channel loss to Bob is $7.5$~dB, resulting in $\eta_B=18\%$.
	The measured $p_{\text{D}}^A$ and $p_{\text{D}}^B$, defined by the raw dark count rate divided by the single count rate, are approximately $5 \times 10^{-4}$ and $2 \times 10^{-3}$, respectively.
	Given the measured parameters, we first simulate the theoretical key-rate bound as a function of two transmitter imperfections: the state visibility $v$, which characterizes the entanglement-state generation, and Alice's heralding efficiency $\eta_A$, which characterizes the photon detection.
	By feeding the calculated probability distribution $p(a,b|A,B)$ associated with varied $\eta_A$ and $v$ to the numerical method, we compute the theoretical key-rate per round and obtain a critical heralding efficiency of 54.6\% and a critical visibility of 79.6\% for a positive key rate as depicted in Fig.~\ref{f3}.
	To evaluate the quality of the generated entanglement states, we conduct state tomography, and the state fidelity is $98$\%.
	Combining with the measured collection efficiency of Alice, we confirm the feasibility of TDI protocol using the present setup.
	
	We acquire $10^7$ rounds of correlation tests for each of the four input combinations, and obtain the full probability distribution of 24 correlation probabilities.
	By taking 14 independent probabilities (see Appendix) into the numerical algorithm, we obtain a final asymptotically secure key rate of 0.048~bit per round, which is denoted by red pentagram in Fig.~\ref{f3}.
	
\begin{figure}[h]
	\centering
	\includegraphics[width=\columnwidth]{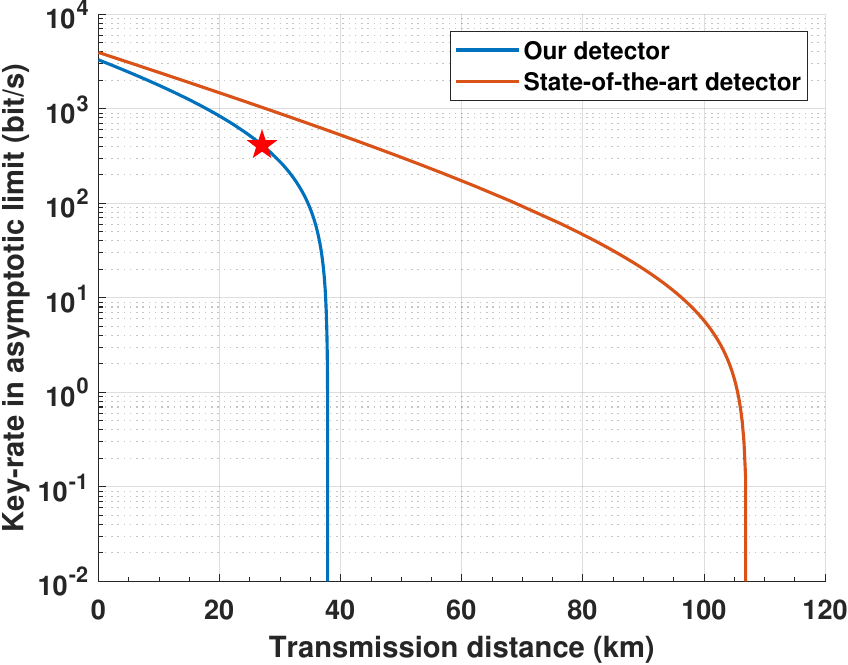}
	\caption{Comparison of our detector and state-of-the-art detector at the receiver with respect to the relation between asymptotically secure key rate and transmission distance from the transmitter to the trusted receiver.
		The spool fiber has a typical loss coefficient of 0.158~dB/km with an estimated polarization-mode-dispersion parameter of 0.021~ps/$\sqrt{\text{km}}$.
		The red pentagram denotes the estimated asymptotically secure key rate based on our experimental data point.
	}\label{f4}
\end{figure}

	If Alice and Bob choose their inputs with equal probability, the estimated asymptotic key rate at a 27-km transmission distance is about 410~bps, see Fig.~\ref{f4}.  
	Assuming $p_\text{D}^B$ remains negligible over a longer distance, the minimum tolerable $\eta_B$ is estimated to be 12\% with the present setup, corresponding to a 38-km fiber transmission.
	Moreover, we numerically simulate the transmission distance assuming state-of-the-art single-photon detectors at the receiver (that is, of 0.02~Hz dark count rate~\cite{liuExperimentalTwinFieldQuantum2023} compared with 50~Hz of ours), which can be improved up to 107~km (denoted by red curve in Fig.~\ref{f4}).
	We note that one can further improve the key-rate performance by experimentally developing highly efficient optical components and single photon detectors as shown in the seminal photonic DI-QKD work~\cite{liuPhotonicDemonstrationDeviceIndependent2022}, and theoretically by increasing the weight of key-generation inputs and using data processing techniques such as noisy preprocessing~\cite{hoNoisyPreprocessingFacilitates2020} and random postselection~\cite{xuDeviceIndependentQuantumKey2022} in practical implementation.

	\section{Discussion}
	To reduce experimental complexity, we did not implement fast basis switching to fully close the signaling loophole and random number generator to close the freedom-of-choice loophole, both of which are crucial to guarantee a positive secret key rate in the practical implementation.
	Moreover, the finite-size effect is another critical ingredient for a practically secure key rate.
	While a comprehensive finite-key analysis for TDI-QKD has yet to be developed, recent advances in finite-size security bounds for 1sDI-QKD based on the entropy accumulation theorem (EAT)~\cite{arnon-friedmanPracticalDeviceindependentQuantum2018,dupuisEntropyAccumulation2020} suggest a promising route toward a finite-key analysis in the TDI-QKD setting~\cite{bluhmComputingKeyRates2026}.
	
	The fair-sampling assumption on Bob's measurement device, though it eases the detection efficiency requirements on the transmitter, poses real security risks.
	There are two ways toward constructing trustworthy detectors for the receiver.
	The first is to admit that the strong fair-sampling assumption is unrealistic and adopt the approximate fair-sampling assumption~\cite{orsucci2020HowPostselection}.
	Under the approximate assumption, a lossy measurement device can be modeled as an approximately unbiased filter followed by an ideal detector.
	Practically, the approximate fair-sampling assumption can be validated by (i) precalibrating the detection efficiency using trusted quantum sources; (ii) publicizing the data and quantifying the security impact of any approximation error.
	The second route is to combine the ideas of semi-quantum game~\cite{buscemiAllEntangledQuantum2012} and self-testing~\cite{supicSelftestingQuantumSystems2020} to circumvent the potential detection loophole in receiver's device.
	However, this approach would require a substantially more sophisticated setup and near-perfect state fidelity~\cite{zhaoDeviceindependentVerificationEinstein2023}. 
	We note that recent progress in improving the detection efficiency in integrated quantum photonics could provide feasible solutions~\cite{liSurpassing99Detection2025}.
	
	The proposed TDI-QKD removes transmitter-side trust requirements without sacrificing the key-rate performance and is compatible with MDI protocol or more precisely the entanglement swapping protocol~\cite{wangHighrateCrosschannelEntanglement2025}.
	The hybrid protocols are expected to achieve longer transmission distance.
	
	
	\section*{Acknowledgments}
	Q.Z. thanks Michele Masani and Erik Woodhead for helpful discussion and their works on developing the algorithm.
	This work was supported by the Innovation Program for Quantum Science and Technology (Grant No. 2024ZD0302500) and Beijing Academy of Quantum Information Sciences (Grant No. Y25NG252702).
	Q.Z. acknowledges support from Beijing Natural Science Foundation under Grant No. Z240007.
	A.M. acknowledges funding from the European Union’s Horizon Europe research and innovation programme under the project "Quantum Security Networks Partnership" (QSNP, Grant Agreement No. 101114043),  from the F.R.S-FNRS through the PDR T.0171.22, from the FWO and F.R.S.-FNRS under the Excellence of Science (EOS) programme project 40007526, from the FWO through the BeQuNet SBO project S008323N.
	
	Funded by the European Union. Views and opinions expressed are, however, those of the author(s) only and do not necessarily reflect those of the European Union. Neither the European Union nor the granting authority can be held responsible for them.
	
	\section*{Data availability}
	The data that support the findings of this article are not publicly available. The data are available from the authors upon reasonable request.

\appendix


\section{The attack model exploiting the signaling loophole}
\begin{figure}[t!]
	\centering
	\includegraphics[width=\columnwidth]{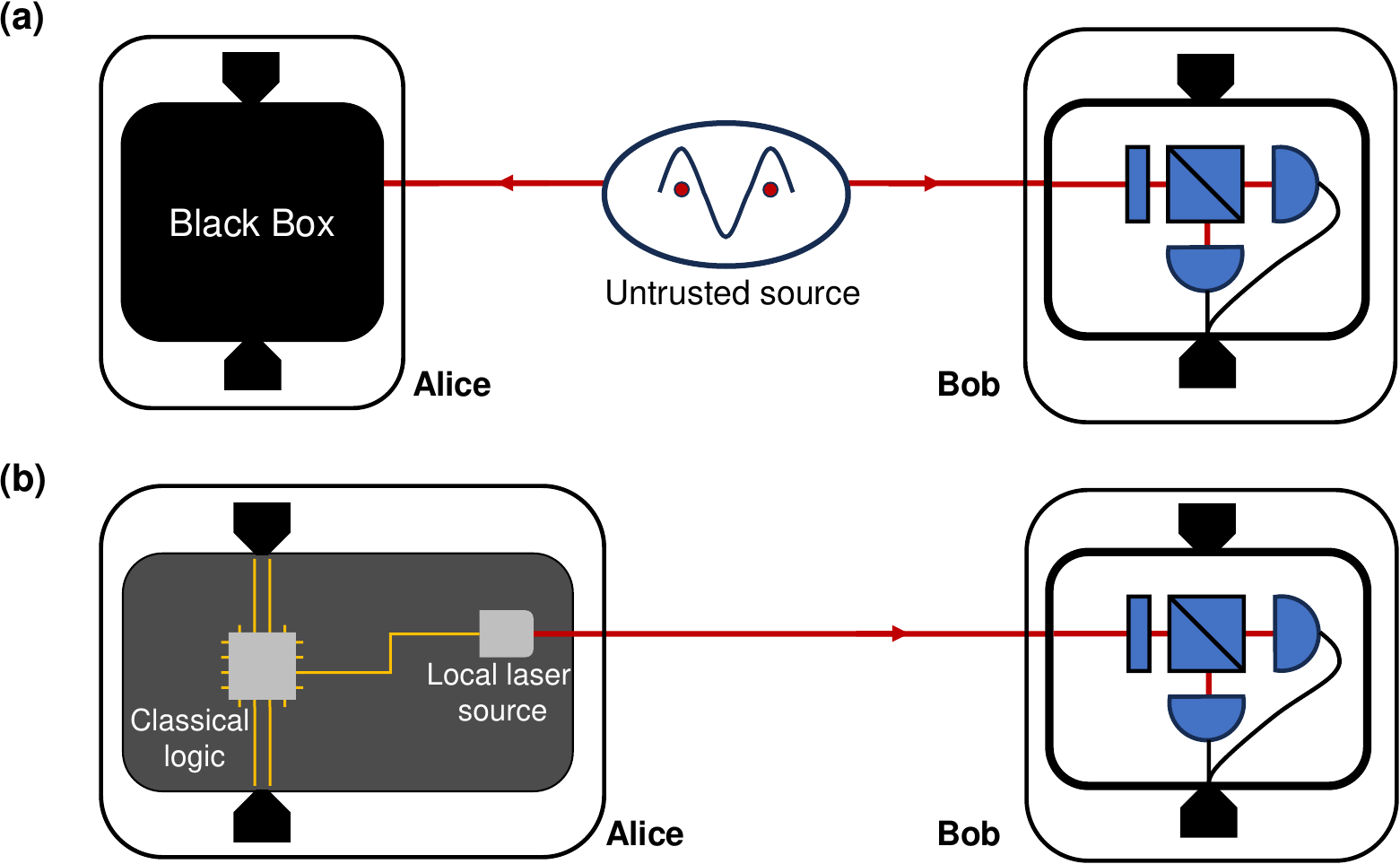}
	\caption{(a) The schematic of 1sDI-QKD in conventional tripartite architecture and (b) the attack model if 1sDI protocol adopts transmitter-receiver architecture.
	}\label{FS0}
\end{figure}

In this appendix, we illustrate the conventional 1sDI-QKD protocol in a tripartite architecture and explain how a signaling loophole can arise when the 1sDI framework is implemented in a transmitter-receiver architecture.
As shown in Fig.~\ref{FS0}(a), an untrusted entanglement source distributes photon pairs to Alice and Bob, who are located in separate, isolated laboratories. 
Each party has access to a trusted source of local randomness and secure classical memory. 
Here, Alice's measurement device is uncharacterized, whereas Bob's measurement device is trusted and characterized. 
Because Alice's laboratory is isolated, her input information is inaccessible to the untrusted entanglement source and to Bob's laboratory before the relevant communication session is completed. 
Consequently, no signaling loophole arises in this conventional tripartite architecture.

It is worth emphasizing that strict spacelike separation is not essential in the cryptographic setting. 
Indeed, cryptographic security imposes a distinct requirement: in addition to ensuring that the input information remains inaccessible to the untrusted source, it requires that measurement outcomes and other security-relevant information remain confined to their respective secure locations. 
If information carrying Alice’s input or measurement outcome were allowed to propagate to another location accessible to an adversary, the security of the protocol could be compromised, even if the relevant events were arranged to be spacelike separated.
Thus, in the cryptographic context, physical isolation of the laboratories \textit{per se} can enforce the required no-signaling condition, without requiring the relativistic no-signaling condition of spacelike separation~\cite{nadlingerExperimentalQuantumKey2022}. 
As long as the relevant local information remains confined to its designated secure laboratory, it cannot be accessed by a remote untrusted device. 
Consequently, the untrusted entanglement source may in principle be located arbitrarily close to Alice, provided that Alice's input and output information remain securely isolated from it.

This differs from the requirements for a loophole-free quantum-steering certification. 
In the latter case, spacelike separation, or an equivalent causal arrangement, is required to enforce the relativistic no-signaling condition independently of assumptions about the physical isolation of the laboratories.
This explains why the abstract parties in Fig.~\ref{f1}(a) are not indicated as in isolated locations.

The situation changes when the untrusted entanglement source is incorporated into Alice's laboratory and effectively treated as part of the same device as her measurement module. 
As depicted in Fig.~\ref{FS0}(b), the combined transmitter could contain arbitrary internal communication between the source and the measurement module. 
In the extreme case, the purported entanglement source and measurement apparatus could instead be replaced by a classical logic unit and a local laser source that coordinate their behavior to reproduce the observed input-output statistics. 
Since the source can then obtain Alice's input before or in real time preparing the state sent to Bob, the emitted state may depend explicitly on Alice's measurement choice. 
The observed statistics can consequently be generated without a genuine quantum assemblage, making an apparent exclusion of local-hidden-state (LHS) models meaningless from the standpoint of transmitter independence and thus invalidating the corresponding security argument.

\begin{table*}[t]
	\caption{POVMs of detection events, considering dark counts and categorized by outcome type.}
	\label{tab:photon_outcomes}
	\centering
	\begin{tabular}{l|l|l|l}
		\hline
		\rule{0pt}{3ex} 
		\textbf{~~~~~~~} & \textbf{Event} & \textbf{POVM} & \textbf{Outcome Type} \\
		\hline
		\rule{0pt}{3ex} 
		~ & First detector sparks, no dark count at second & $\eta \times (1 - p_\text{D}) \times P_1 $ & Single click in the first \\
		\multirow{2}*{Photon received} & Second detector sparks, no dark count at first & $ \eta \times (1 - p_\text{D}) \times P_2 $ & Single click in the second \\
		~ & Photon at first detector, dark count at second & $\eta \times p_\text{D} \times P_1 $ & Double click \\
		~ & Photon at second detector, dark count at first & $\eta \times p_\text{D} \times P_2 $ & Double click \\
		\hline
		\hline
		\rule{0pt}{3ex} 
		~ & No dark count in either detector & $(1-\eta)\times (1 - p_\text{D})^2 \times \mathbbm{1} $ & No click \\
		\multirow{2}*{Photon lost} & Dark count in the first detector only & $(1-\eta)\times p_\text{D} \times (1 - p_\text{D}) \times \mathbbm{1} $ & Single click in the first \\
		~ & Dark count in the second detector only & $(1-\eta)\times (1 - p_\text{D}) \times p_\text{D} \times \mathbbm{1} $ & Single click in the second \\
		~ & Dark counts in both detectors & $(1-\eta)\times p_\text{D}^2 \times \mathbbm{1} $ & Double click \\
		\hline
		
	\end{tabular}
	
\end{table*}

\section{The detection model}
In this appendix, we clarify in detail the model of the POVM set of Alice and Bob, taking into account the photon loss and dark counts. 
Let us first consider the case where the dark count is zero.
In the experiment, when a photon arrives at the measurement device (with probability $\eta$), it encounters a polarizing beam splitter, which then directs it to one of two detectors based on its polarization state. 
If the photon is lost (with probability $1-\eta$), then neither detector clicks.

Let us now consider that the detectors of Alice and Bob have dark counts, with each of the two possible outcomes having a probability of $p_\text{D}$ of registering a spurious click in every round of the protocol. 
Our setup does not allow us to differentiate whether a click event was due to a genuine photon detection, or a dark count.
We can delineate the four possible detection outcomes as follows:
\begin{enumerate}
	\item \textbf{Single click in the first detector:} This outcome indicates that only the detector "1" clicked. 
	It could arise from a genuine photon detection or a dark count.
	\item \textbf{Single click in the second detector:} This outcome indicates that only the detector "2" clicked, which can result from a photon in an orthogonal state with respect to outcome "1" was detected, or a dark count.
	\item \textbf{Double click:} It occurs when both detectors register a click simultaneously. This can arise from both detectors experiencing dark counts, or one detecting a photon while the other registers a dark count.
	\item \textbf{No click:} When neither detector registers a click, suggesting the photon was lost and no dark counts occurred.
\end{enumerate}

Let us now evaluate the POVMs related to each of these four events. 
We denote the ideal POVMs by $P_{1}$ and $P_{2}$ of the photon being detected by the first and second detectors, respectively, and $\eta$ the efficiency of the photon's arrival at the detectors, and $p_\text{D}$ the probability of a dark count occurring in each detector.
Assuming a photon is indeed received by the measurement device, four possible events can occur. 
We list them in the upper half of Table~\ref{tab:photon_outcomes}.
For the case where the photons are lost, we have four cases which are listed in the lower half of the table.

To compute the measurement operators of Alice, given the measurement projectors $P_{1|A}$ for outcome "1" and $P_{2|A}$ for outcome "2", we need to consider the probabilities of each event as previously discussed. 
We obtain in this way four measurement operators $P_{a|A}(\eta_A,p_\text{D})$ corresponding to the four outcome types, where $\eta_A$ is Alice's detection efficiency.

\begin{align}
	P_{1|A}(\eta_A,p_\text{D}) &= \eta_A(1-p_\text{D}) 			\nonumber			P_{1|A}+(1-\eta_A)p_\text{D}(1-p_\text{D})\mathbbm{1}, \\
	P_{2|A}(\eta_A,p_\text{D}) &= \eta_A(1-p_\text{D}) 	\nonumber	P_{2|A}+(1-\eta_A)p_\text{D}(1-p_\text{D})\mathbbm{1}, \\ \nonumber
	P_{\text{DC}|A}(\eta_A,p_\text{D}) &= \big(p_\text{D}\eta_A +(1-\eta_A)p_\text{D}^2\big)\mathbbm{1}, \\	
	P_{\text{NC}|A}(\eta_A,p_\text{D}) &= (1-\eta_A)(1-p_\text{D})^2\mathbbm{1}. \nonumber
\end{align} 
For simplicity, we group Alice's no clicks and double clicks into a single outcome, expressing as
\begin{equation}
	P_{\varnothing|A}(\eta_A,p_\text{D}) = \big(p_\text{D}\eta_A	\nonumber +(1-\eta_A)(p_\text{D}^2+(1-p_\text{D})^2)\big)\mathbbm{1}.		
\end{equation}

As for Bob's measurements, his ideal POVMs are denoted by $P_{b|B}$. 
Given the assumption that the probability of a click in Bob's detector is independent of the basis choice, we are allowed to discard events where both his detectors do not click or where they both click. 
As we discard these events, we need to renormalize Bob's POVMs dividing them by the probability of having only one click. 
Consequently, the measurement operators of Bob $P_{b|B}(\eta_B,p_\text{D})$ can be expressed in terms of the ideal projectors $P_{1|B}$ as
\begin{equation}
	P_{1|B}(\eta_B,p_\text{D}) = \frac{ \eta_B(1-p_\text{D}) P_{1|B} +(1-\eta_B)p_\text{D}(1-p_\text{D})\mathbbm{1}}{1-p_\text{D}\eta_B -(1-\eta_B)(p_\text{D}^2+(1-p_\text{D})^2)},	\nonumber
\end{equation} and $P_{2|B}(\eta_B,p_\text{D})=\mathbbm{1}-P_{1|B}(\eta_B,p_\text{D})$. 
Here, $\eta_B$ is the detection efficiency on Bob's side.

\section{Key-rate lower bounds by the conditional von Neumann entropy}
We evaluate the key rate in the asymptotic regime under one-way classical postprocessing from Alice to Bob using the Devetak-Winter bound:
\begin{equation}
	r_{\infty} \ge H(b|E,B_1)-H(b|a,A_1,B_1),
\end{equation}
where $H(b|E,B_1)$ represents the conditional von Neumann entropies of Bob's outcomes by measurement $B_1$ given Eve’s information, and $H(b|a,A_1,B_1)$ represents the conditional Shannon entropies of outcomes of $B_1$ given Alice's outcome by $A_1$.
The value of $H(b|a,A_1,B_1)$ can be directly inferred from the input-output statistics obtained %
from measurements of $A_1$ and $B_1$:
\begin{equation}
	H(b|a,A_1,B_1) = \sum_{a,b} p(a,b|A_1,B_1)\log_2\frac{p(a,b|A_1,B_1)}{p(a|A_1)}.
\end{equation}

In contrast, estimating the value of $H(b|E,B_1)$ presents a considerably greater challenge. 
It is well established that utilizing full input-output statistics, rather than specific combinations of correlators such as Bell inequalities, yields higher secrecy rates~\cite{schwonnekDeviceindependentQuantumKey2021}. 
Recent advances have demonstrated that tighter lower bounds on the key rates of DI protocols can be derived by formulating the conditional von Neumann entropy of composite quantum systems as a sequence of optimization problems~\cite{brownComputingConditionalEntropies2021,brownDeviceindependentLowerBounds2024}. 
This advantage of full input-output statistics over steering inequalities has also been proven in the one-sided device-independent (1sDI) scenario~\cite{branciardOnesidedDeviceindependentQuantum2012}. 
In particular, a minimal detection-efficiency threshold allowed by quantum incompatibility theory~\cite{acinNecessaryDetectionEfficiencies2016,masiniJointmeasurabilityQuantumCommunication2024} is achieved using the newly developed numerical techniques~\cite{masiniOnesidedDIQKDSecure2026}.
Moreover, these numerical techniques do not involve post-selection, which enables a comparatively simple extension of security proofs against coherent attacks as well as finite-size effects by employing the entropy accumulation theorem (EAT)~\cite{arnon-friedmanPracticalDeviceindependentQuantum2018,dupuisEntropyAccumulation2020}.

This appendix outlines the framework of non-commutative polynomial optimization employed to establish lower bounds on the conditional von Neumann entropy $H(B_1|E)$. 
This methodology is fundamental to the security validation of the TDI protocol discussed in the main text. 
For a comprehensive derivation regarding the approximation of the conditional von Neumann entropy, we refer the reader to the original work cited in Ref.~\cite{brownDeviceindependentLowerBounds2024}.

Our approach utilizes the Gauss-Radau quadrature, a variant of Gaussian quadrature designed for numerical integration. 
This method employs $m+1$ points to exactly integrate polynomials of degree $2m-1$, effectively fitting all polynomials up to degree $2m$. 
Specifically, we implement an $m$-point Gauss-Radau quadrature on the interval $(0,1]$ with a weighting function $W(x)=1$, fixing the endpoint $t_m=1$. 
The corresponding nodes $t_1,\dots,t_m$ and weights $w_1,\dots,w_m$ are computed efficiently using Legendre polynomials. 
We define the coefficients $\alpha_i$ as:
$$ \alpha_i=\frac{3}{2}\max \left\{ \frac{1}{t_i},\frac{1}{1-t_i}\right\}. $$

Using these parameters, a lower bound for the conditional entropy $H(b|E,B_1)$ is given by:
$$ H(b|E,B_1) \geq \sum_{i=1}^{m-1}\frac{w_i}{t_i \ln(2)}+ \sum_{i=1}^{m-1}\frac{w_i}{t_i \ln(2)}O_i, $$
where $O_i$ represents the solution to the following non-commutative polynomial optimization problem:

\begin{widetext}
	\begin{align} 
		\text{inf}\quad & \sum_b \langle{\psi}|P_{b|B_1}(Z_{b,i}+Z^*_{b,i}+(1-t_i)Z_{b,i}^*Z_{b,i})+t_iZ_{b,i}Z^*_{b,i}|{\psi}\rangle \nonumber\\ 
		\text{s.t.}\quad & \langle{\psi}|P_{a|A}P_{b|B}|{\psi}\rangle=p(a,b|A,B), \nonumber\\ 
		& \sum_a P_{a|A}=\sum_b P_{b|B}=\mathbbm{1}, \nonumber\\ 
		& P_{a|A}\geq 0,\quad P_{b|B}\geq 0, \nonumber\\ 
		& Z_{b,i}^*Z_{b,i}\leq\alpha_i,\quad Z_{b,i}Z^*_{b,i}\leq\alpha_i, \nonumber\\ 
		& [P_{b|B},P_{a|A}]=[P_{b|B},Z^{(*)}_{b,i}]=[Z_{b,i}^{(*)},P_{a|A}]=0, \nonumber\\ 
		& \{B_1, B_2\} = 0. \label{eq:constr-anti} 
	\end{align}
\end{widetext}

In this formulation, $Z_n$ denotes bounded operators. 
For the final constraint Equation~\eqref{eq:constr-anti}, we enforce the anti-commutation relation between the observables $B_1$ and $B_2$, where $B_i=P_{1|B_i}-P_{2|B_i}$, to achieve a better key-rate performance.
The validity of this constraint is rooted in the trust assumption placed on the receiver's devices within the 1sDI framework, which in practice necessitates careful calibration of Bob's measurement operations.
In our experimental demonstration, this constraint is realized by corresponding $B_1$ to $\sigma_\textsf{Z}$ and $B_2$ to $\sigma_\textsf{X}$.
In specific, $\sigma_\textsf{Z}$ is implemented via a half-wave-plate at 0$^{\circ}$ followed by a polarization-beam-splitter, while $\sigma_\textsf{X}$ is implemented via a half-wave-plate rotated by 22.5$^{\circ}$.
We note that this anti-commutation constraint is not essential for a positive key rate.
As shown in Ref.~\cite{masiniOnesidedDIQKDSecure2026}, the numerical analysis produces a positive key rate without the anti-commutation constraint when the detection efficiency is sufficiently high ($\ge 70\%$).

The probabilities $p(a,b|A,B)$ are determined experimentally.
In the presence of noisy preprocessing, the measurement operators $P_{a|1}$ in the objective function are replaced according to:
\begin{align} 
	P_{1|B_1} &\mapsto (1-q)P_{1|B_1}+qP_{2|B_1}, \nonumber\\ 
	P_{2|B_1} &\mapsto (1-q)P_{2|B_1}+qP_{1|B_1}. \nonumber
\end{align} 
This optimization problem is relaxed into a hierarchy of semi-definite programs (SDPs) following the methodology of Ref.~\cite{navascues2008ConvergentHierarchy}.
The tightness of the resulting lower bound improves with higher levels of the hierarchy and an increased number of quadrature points. 
For the computations presented in this work, particularly in scenarios assuming anti-commuting observables for Bob, we fixed the level of the localizing matrices in the NPA hierarchy to $1$ and utilized $m=20$ quadrature points. 
To ensure consistency, the level of the principal moment matrix was set slightly higher to encompass all terms present in the localizing matrices.

Notably, our analysis indicated that incorporating localizing matrices for the constraints $Z_{b,i}^* Z_{b,i}\leq\alpha_i$ and $Z_{b,i}Z_{b,i}^*\leq\alpha_i$ did not enhance the bounds. 
Consequently, we adopted the simpler, linear constraints $\langle{\psi}|Z_{b,i}^*Z_{b,i}|{\psi}\rangle\leq\alpha_i$ and $\langle{\psi}|Z_{b,i}Z_{b,i}^*|{\psi}\rangle\leq\alpha_i$, which remain valid for establishing a lower bound.

In the analysis of the TDI protocol under the fair sampling assumption on Bob's side, we omitted localizing matrices entirely. 
This decision was based on the absence of an anti-commutativity requirement and the sufficiency of the aforementioned linear constraints. 
However, achieving convergence in this scenario necessitated employing a principal moment matrix at level $2+ABZ$.

\begin{figure}[h!]
	\centering
	\includegraphics[width=\columnwidth]{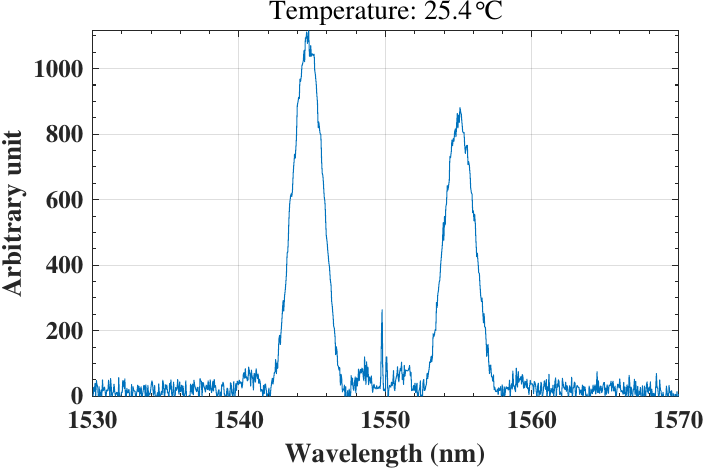}
	\caption{Normalized spectra of the SPDC photon pairs.	
	}\label{FS1}
\end{figure}

\section{Experimental details}
A continuous-wave (cw) pump centered at 775~nm with a 2.5~kHz linewidth (Haomin Opt.) 
is directed into the Sagnac-loop interferometer by a dichroic mirror (DM).
Inside the loop, a periodically poled lithium niobate (PPLN) crystal is pumped in both counter-propagating directions, generating polarization-entangled photon pairs via spontaneous parametric down-conversion.
The PPLN crystal dimensions are: $2\times1\times10$~mm$^3$ (W$\times$H$\times$L).
The idler photons are sent to the photon storage module which comprises a 30-m fiber and a polarization controller set, then directed to Alice's detection module.
The active half-wave plate (a-HWP) is to implement the input settings for Alice.
In specific, HWP 0$^{\circ}$ for implementing $\sigma_\textsf{Z}$ and HWP 22.5$^{\circ}$ for implementing $\sigma_\textsf{X}$.
Similar configuration applies to the measurements at the receiver.
The idler photons are collected by a 400~mm lens and coupled into single-mode fibers using an 11~mm collimator~\cite{schwallerOptimizingCouplingEfficiency2022}.
The signal photons are sent to Bob through a 27-km ultra-low-loss spool fiber (G654.C ULL) with a typical loss coefficient of 0.158~dB/km.
The combination of an a-HWP and two active quarter-wave plates (a-QWPs) is for the polarization compensation.

We use a spectrometer to characterize the wavelength distribution of the SPDC photon pairs.
The normalized spectra are presented in Fig.~\ref{FS1}, at a temperature of 25.4$^\circ$C.
The employed PPLN crystal has poling period of 9.02~$\mu$m for a type-II quasi-phase-matching (QPM).
The linewidth of the photon pairs is also characterized as $~$2.4~nm, which is beneficial for a better robustness against decoherence compared with type-0 QPM~\cite{steinlechner2014EfficientHeralding}.

Photons are detected using superconducting nanowire single-photon detectors (SNSPDs) with approximately 90\% detection efficiency and a raw dark-count rate of approximately 50~Hz. 
Filtering elements suppressing residual pump light and background noise are omitted for clarity. 
The total collection efficiency in Alice's arm is calibrated to be 61\% (2.15~dB), including losses from bulk optics (0.35~dB), spherical aberration (0.25~dB), fiber components (0.2~dB), space-to-fiber coupling (0.79~dB), correlated-mode coupling (0.06~dB), and detector inefficiency (0.5~dB).
The total collection efficiency in Bob's arm is calibrated to be 18\% (7.5~dB), including losses from bulk optics (0.4~dB), spherical aberration (0.25~dB), spool fiber (4.3~dB), fiber components (0.6~dB), space-to-fiber coupling (1.19~dB), correlated-mode coupling (0.06~dB), and detector inefficiency (0.7~dB).
Photon clicks at the four outputs, detected by the SNSPDs, are registered by a time tagger (not shown in the schematic), from which we extract the coincidence events between the two parties.
The null results of the transmitter are extracted by excluding the coincidence events from the single channel detection events of the receiver.
This method is also used to estimate the detection efficiency at the transmitter, known as heralding efficiency or Klyshko efficiency~\cite{klyshkoUseTwophotonLight1980a} defined as the coincidence counts divided by the single counts of the opposite arm.

\begin{figure}[h]
	\centering
	\includegraphics[width=\columnwidth]{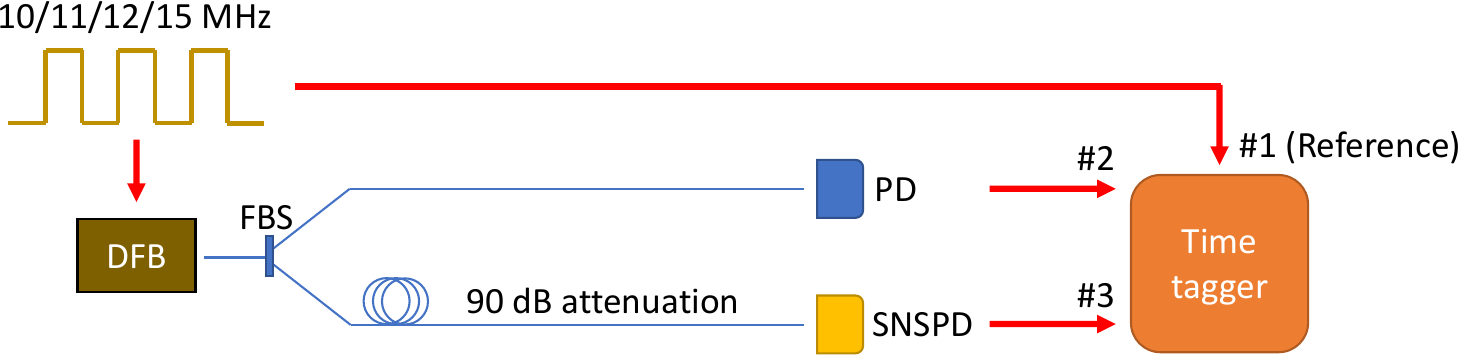}
	\caption{SNSPD delay test schematic.	
	}\label{FS4}
\end{figure}

\begin{figure*} [t!]
	\centering
	\subfloat[\label{fig:5a}]{
		\includegraphics[width=\columnwidth]{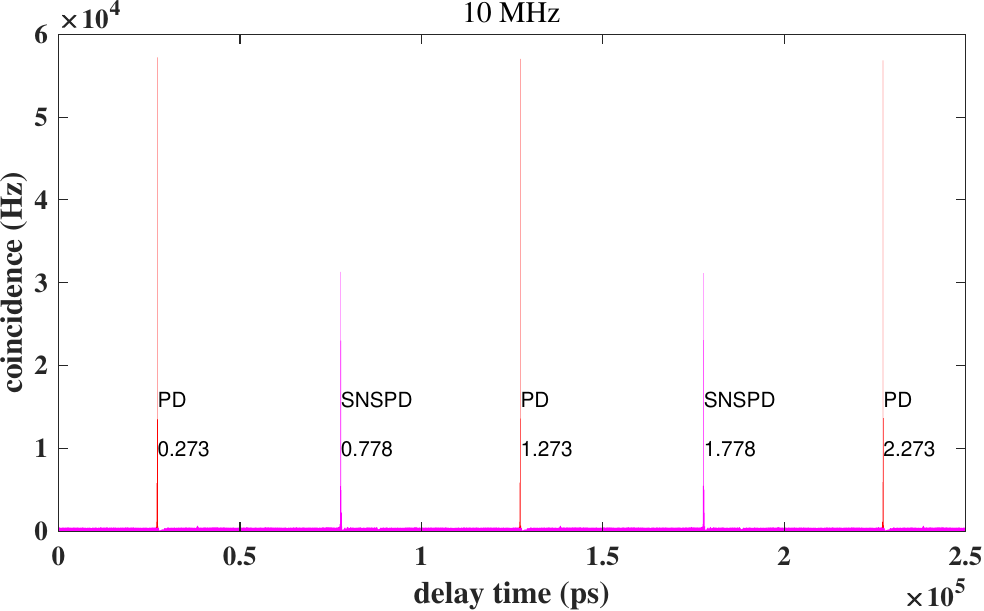}}
	\subfloat[\label{fig:5b}]{
		\includegraphics[width=\columnwidth]{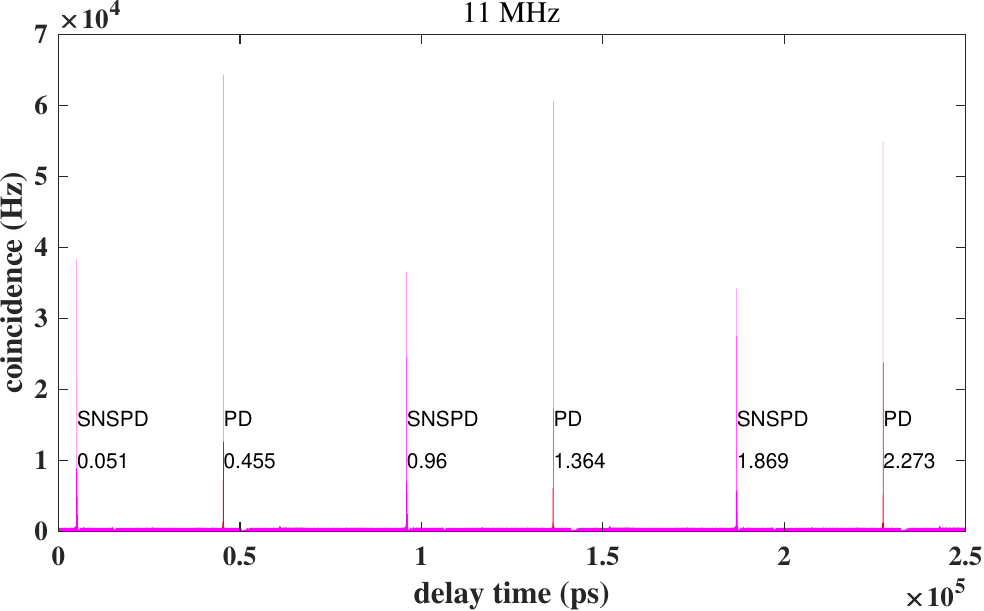}}\\
	\subfloat[\label{fig:5c}]{
		\includegraphics[width=\columnwidth]{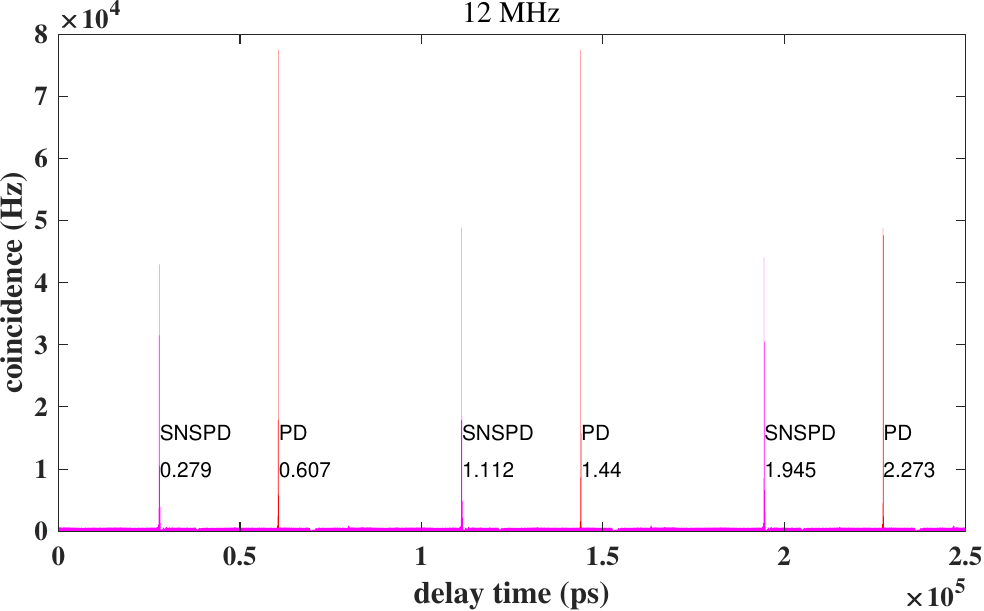}}
	\subfloat[\label{fig:5d}]{
		\includegraphics[width=\columnwidth]{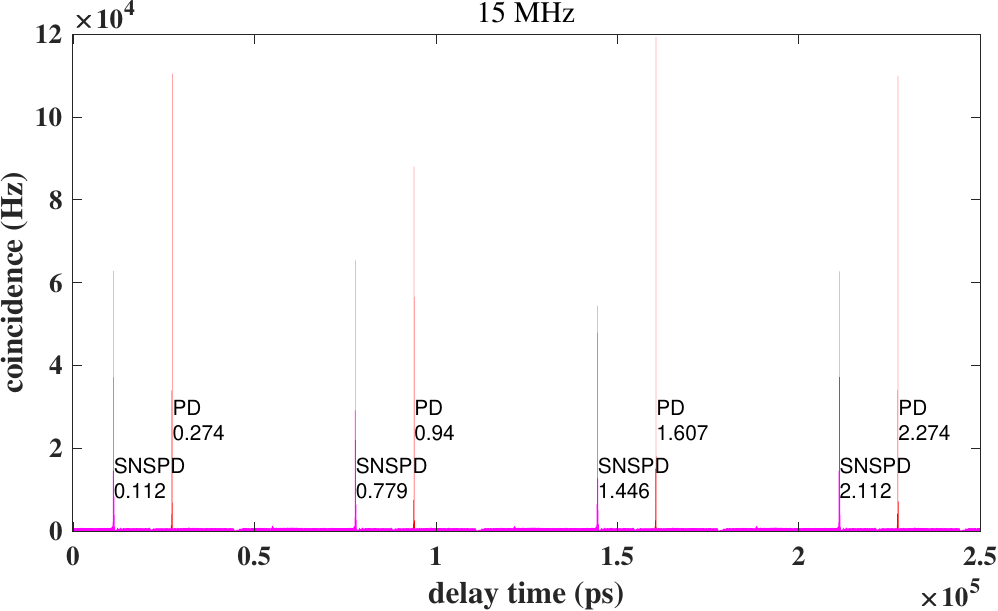}}
	\caption{SNSPD delay test results.}
	\label{FS5} 
\end{figure*}

The measurement interval of the SNSPD is calibrated by comparing the detection time-tag of the SNSPD with the detection time-tag of a high-speed photodiode (PD) of 10~GHz bandwidth.
As shown in Fig.~\ref{FS4}, we employ a distributed feedback laser (DFB) to generate pulsed laser to impact the PD and record the time difference between the PD signal and the reference signal.
Then, the same fiber channel is attenuated to a single-photon level to impact the SNSPD, and we record the time difference between the signals.
We using four adjacent repetition-frequency of 10, 11, 12, and 15~MHz to eliminate the potential multi-period error.
The results are shown in Fig.~\ref{FS5}.
We confirm that the measurement interval of the SNSPD is about 50~ns.

\begin{figure}[h!]
	\centering
	\includegraphics[width=\columnwidth]{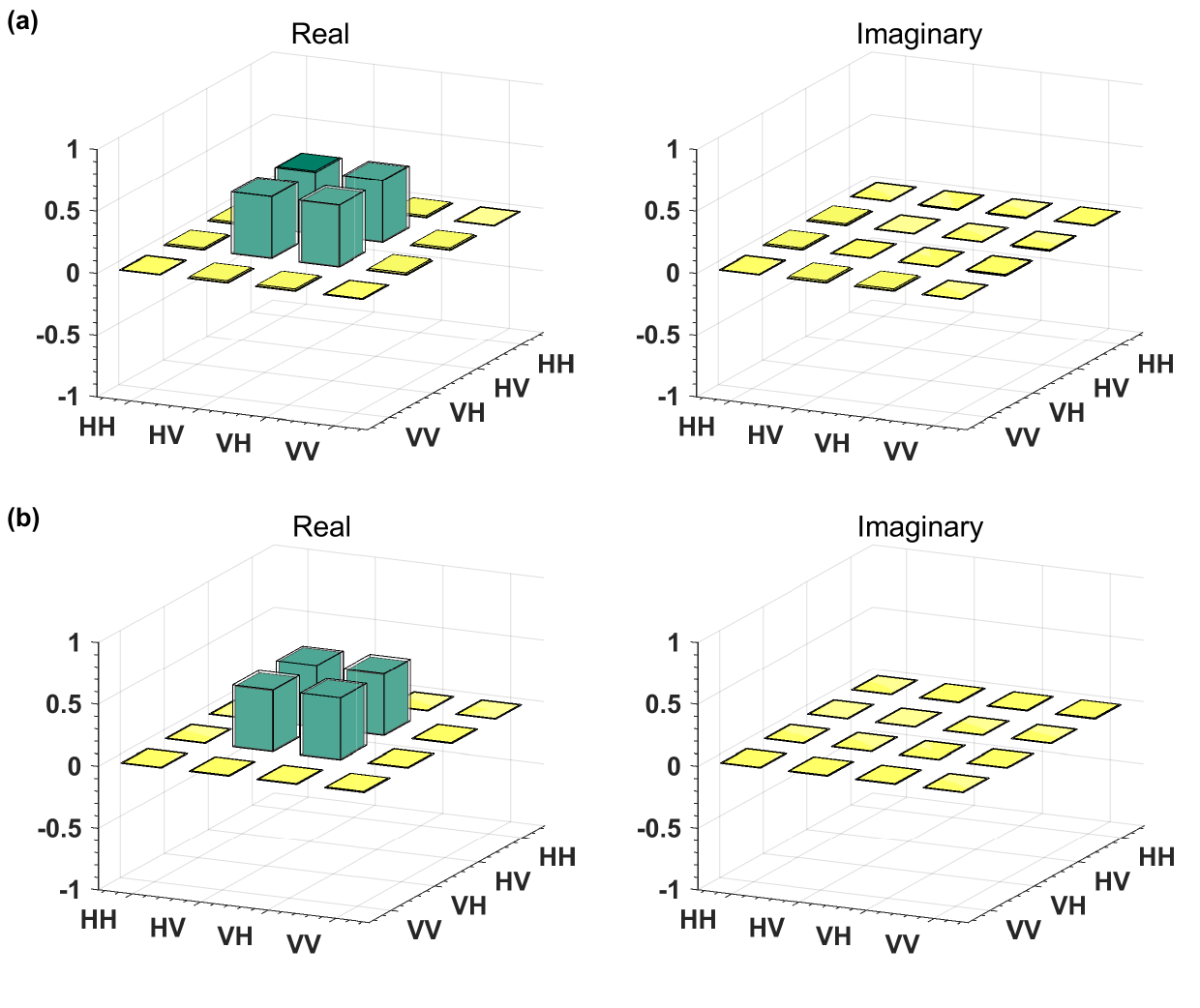}
	\caption{State tomography results. 
		(a) Reconstructed state density matrix from the experimental data without applying spool fiber.
		(b) Reconstructed state density matrix with the 27-km spool fiber transmission.
		The state density matrix of the maximally entangled state $\left(\kt{01}+\kt{10}\right)/ \sqrt{2}$ are denoted by transparent bars for reference. 	
	}\label{FS2}
\end{figure}

\section{State tomography and noise analysis}

State tomography was performed using an overcomplete measurement scheme~\cite{agnewTomographyQuantumState2011}, with measurements in the $\sigma_\textsf{X}$, $\sigma_\textsf{Y}$, and $\sigma_\textsf{Z}$ bases, yielding six measurement outcomes for each qubit. The reconstructed state exhibited a visibility of $99.25(3)\%$ before transmission through the 27-km fiber spool, which decreased to $97.98(4)\%$ after transmission. The corresponding tomographic results are presented in Fig.~\ref{FS2}.

The deviation from the ideal state can be attributed primarily to (i) detector dark counts and statistical fluctuations associated with photon counting, and (ii) polarization-mode dispersion (PMD) introduced by the fiber spool. In particular, the interference visibility measured at Bob was $0.996$ before and $0.9837$ after fiber transmission. Using the Gaussian decoherence model
$$
V(\Delta T)=\exp\left[-\left(\frac{\Delta T}{T_c}\right)^2\right],
$$
where $T_c$ denotes the coherence time, we estimate a differential group delay of approximately $0.11$ ps from the observed reduction in visibility. Here, $T_c$ is estimated to be approximately $1$ ps from the measured optical spectrum with a bandwidth of 2.4 nm.

Based on the estimated differential group delay, we obtain a PMD coefficient of approximately $0.021$ ps/$\sqrt{\text{km}}$ for the 27-km ultra-low-loss (ULL) fiber, in good agreement with the manufacturer’s specification of $\leq 0.04$ ps/$\sqrt{\text{km}}$. These results indicate that PMD is a measurable but relatively minor source of state degradation over the 27-km transmission distance. For substantially longer fiber links, however, fiber with a lower PMD coefficient or an appropriate PMD-compensation scheme would be required to maintain sufficiently high state visibility.

\begin{figure}[t]
	\centering
	\includegraphics[width=\columnwidth]{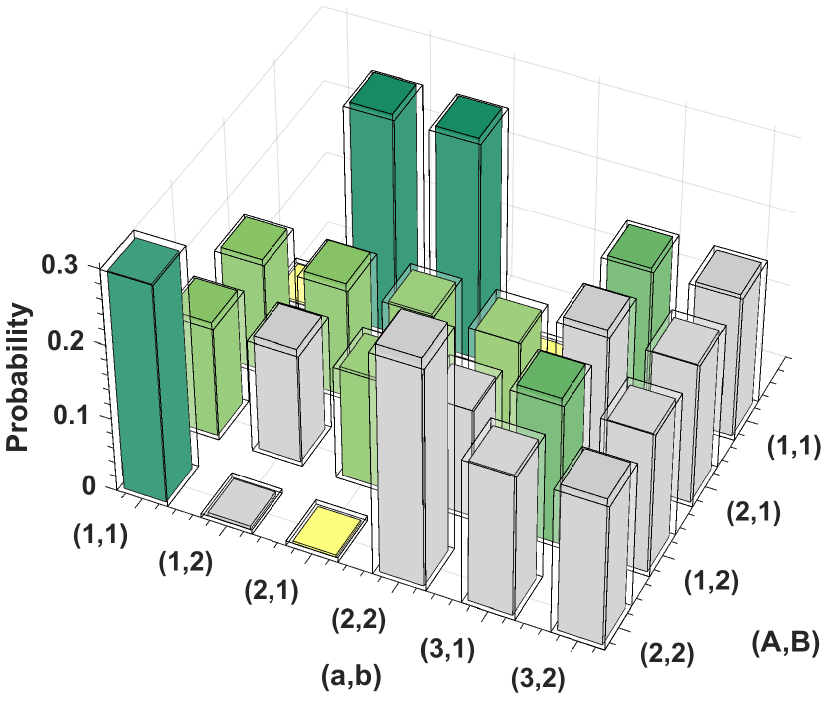}
	\caption{The results of full probability distribution.
		The ten gray bars denote the redundant correlation probabilities due to the probability conservation  and no-signaling constraints. 
	}\label{FS3}
\end{figure}

\section{Full probability distribution of the key-rate data point}

Using the reconstructed density matrix together with the calibrated collection efficiency, we calculate the expected probability distribution over all 24 outcomes, shown as transparent bars in Fig.~\ref{FS3}. Experimentally, we acquire $10^7$ correlation-test rounds for each of the four input combinations, and the corresponding normalized probability distributions are shown as solid bars in Fig.~\ref{FS3}. The experimental results exhibit good agreement with the theoretical predictions.

The probability distribution is subject to normalization constraints, while the no-signaling principle imposes additional linear constraints on the probabilities~\cite{10.1093/oso/9780198788416.001.0001}. Consequently, only 14 of the 24 probabilities are independent. This distinction is important when incorporating the experimental data into the subsequent numerical analysis: imposing redundant constraints, such as those indicated in gray, can overconstrain the optimization problem and render it infeasible, particularly in the presence of experimental statistical fluctuations. We therefore retain only the 14 independent parameters in the analysis to avoid introducing mutually inconsistent constraints.


%

\end{document}